\PassOptionsToPackage{table}{xcolor}
\documentclass[sigconf]{acmart}
\acmConference[ICSE 2024]{46th International Conference on Software Engineering}{April 2024}{Lisbon, Portugal}

\AtBeginDocument{%
  }

\copyrightyear{2024} 
\acmYear{2024} 
\setcopyright{rightsretained} 
\acmConference[ICSE-SEIP '24]{46th International Conference on Software Engineering: Software Engineering in Practice}{April 14--20, 2024}{Lisbon, Portugal}
\acmBooktitle{46th International Conference on Software Engineering: Software Engineering in Practice (ICSE-SEIP '24), April 14--20, 2024, Lisbon, Portugal}\acmDOI{10.1145/3639477.3639744}
\acmISBN{979-8-4007-0501-4/24/04}

\usepackage{amsmath,amsfonts}
\usepackage{algorithmic}
\usepackage{graphicx}
\usepackage{textcomp}
\usepackage{comment}
\usepackage[framemethod=tikz]{mdframed}

\begin{document}

\title{Automated Security Findings Management: A Case Study in Industrial DevOps}

\author{Markus Voggenreiter}
\orcid{0000-0003-3964-1983}
\affiliation{%
  \institution{Siemens Technology, LMU Munich}
  \city{Munich}
  \country{Germany}
}

\author{Florian Angermeir}
\orcid{0000-0001-7903-8236}
\affiliation{%
  \institution{fortiss GmbH}
  \city{Munich}
  \country{Germany}
}

\author{Fabiola Moy\'on}
\orcid{0000-0003-0535-1371}
\affiliation{%
  \institution{Siemens Technology, Technical University of Munich}
  \city{Munich}
  \country{Germany}
}

\author{Ulrich Schöpp}
\orcid{0000-0002-5445-9461}
\affiliation{%
  \institution{fortiss GmbH}
  \city{Munich}
  \country{Germany}
}

\author{Pierre Bonvin}
\email{pierre-louis.bonvin@hm.edu}
\affiliation{%
  \institution{Munich University of Applied Sciences}
  \city{Munich}
  \country{Germany}
}

\renewcommand{\shortauthors}{Voggenreiter et al.}

\begin{abstract}

In recent years, DevOps, the unification of development and operation workflows, has become a trend for the industrial software development lifecycle. Security activities turned into an essential field of application for DevOps principles as they are a fundamental part of secure software development in the industry. A common practice arising from this trend is the automation of security tests that analyze a software product from several perspectives. To effectively improve the security of the analyzed product, the identified security findings must be managed and looped back to the project team for stakeholders to take action. This management must cope with several challenges ranging from low data quality to a consistent prioritization of findings while following DevOps aims. To manage security findings with the same efficiency as other activities in DevOps projects, a methodology for the management of industrial security findings minding DevOps principles is essential.

In this paper, we propose a methodology for the management of security findings in industrial DevOps projects, summarizing our research in this domain and presenting the resulting artifact. As an instance of the methodology, we developed the Security Flama, a semantic knowledge base for the automated management of security findings. To analyze the impact of our methodology on industrial practice, we performed a case study on two DevOps projects of a multinational industrial enterprise. 
The results emphasize the importance of using such an automated methodology in industrial DevOps projects, confirm our approach's usefulness and positive impact on the studied projects, and identify the communication strategy as a crucial factor for usability in practice. 
\end{abstract}



\maketitle

\section{Introduction}
\label{sec:intro}
Over the last few years, the software development strategy in the industry has changed from traditional, waterfall-oriented models to iterative and incremental models. A change in the distribution, operation, and maintenance of industry software further caused many practitioners to follow DevOps principles \cite{Hasselbring_2019}. Amongst multidisciplinary collaboration in the project, one of the key advantages is the automation of tests \cite{Kalliosaari_2016}. 
To comply with industrial security standards and best practices \cite{bsimm12,owasp_samm,safecode,62443,sp800-218}, integrating security aspects in the automation testing strategy is relevant for industrial companies \cite{Moyon_2020}, and represents a commonly recommended measure in software development lifecycles \cite{Khan_2021}. 
Due to the complexity and size of industrial software, the aspects under test range from software and infrastructure code over configurations and third-party dependencies to the running applications and their production environment.  

In practice, problems with the security of the software can be identified in each of its components. Hence, each security test category adds another perspective on the overall picture of the software security status and consequently utilizes a different terminology to describe the flaw. Regardless of the security problems cause, it must be managed and reacted to. For the purposes of this manuscript, we define the term \textit{security finding} as any weakness related to the security of a software product that was identified but not yet confirmed or further processed. This definition is aligned with the ISO security standards treating software vulnerability analysis, which define weakness as software product characteristics that, in proper conditions, could contribute to introducing vulnerabilities \cite{ISO_20004}. Such security findings are detected, e.g., during automated security tests in CI/CD pipelines, manual code reviews, or continuous monitoring activities in production environments. Examples range from hard-coded passwords identified early in the life cycle to publicly known vulnerabilities found during monitoring.
Minding DevOps principles, security findings must be managed in close collaboration between stakeholders, security experts, and the development team with the ability to mitigate or fix them while trying to reduce the size of work packages and minimize lead time. However, the automation of tests often results in huge low-quality data sets, with security-specific terminology acquired at distinct stages of the software engineering life cycle. Dealing with them manually requires substantial time and effort from the entire project team. Consequently, the management of security findings in the industry represents a considerable challenge, especially if it should be performed with the same efficiency as other DevOps practices. 

This paper proposes a methodology for the management of security findings in industrial DevOps projects. We evaluate the impact of the methodology on industrial practice by conducting a case study with two software engineering projects. 
In this context, we developed an instance of the methodology, called the \textbf{Security} \textbf{F}eedback \textbf{L}oop \textbf{A}nalysis and \textbf{M}anagement \textbf{A}pplication (\textit{Security Flama}). The \textit{Security Flama} collects security testing reports from various sources and automatically improves data quality, supporting practitioners to track and take action on security findings. In addition, it communicates evidence-based results tailored toward the individual needs of different stakeholders, such as developers, project managers, or security experts. 
The evaluation provides three main outcomes. First, the ability to continuously visualize the current state of security findings has shown to be beneficial for industrial software projects. Second, our methodology in particular was perceived as highly useful and has noticeably impacted the amount of security findings in one of the projects.  Finally, the communication strategy with the project team and stakeholders was identified as a key factor in looping back security findings information from testing to initial DevOps phases like plan and code. 
In summary, both the methodology and its instance, the \textit{Security Flama} implement the \textit{security feedback loop}, a concept we define as applying the DevOps principle of continuous feedback \cite{Kimphoenix_2018} for the use case of security findings in the software product.

\textbf{Contributions}
With this paper, we make the following contributions: 
1. A methodology for the automated management of security findings in industrial software engineering projects following DevOps principles. 
2. Guidance on how to instantiate the methodology, including its key features, in order to replicate the \textit{Security Flama}\footnote{The reader may understand that the Security Flama source code is protected by a non-disclosure agreement with the industry partner.} for industrial usage.
3. An industry case study on the potential of the methodology gathering quantitative and qualitative data simultaneously on the management of security findings. 
\section{Background and Related Work}
\label{sec:back}
To foster productive collaboration between team members with varying backgrounds, organizations turned towards DevOps to bridge the gap between development and operations \cite{Hasselbring_2019, Plant_2021} using cross-functional teams \cite{Wiedemann_2019, Fitzgerald_2017}. However, neither academia nor industry has a universal definition for DevOps and the principles that should be followed \cite{Fitzgerald_2017}. Humble and Molesky propose four core principles, when bridging the gap between development and operations \cite{Humble_2011}. Similarly, Gene Kim describes the fundamental principles of DevOps as the ''three ways'' \cite{Kimphoenix_2018}, covering parts of the concepts provided by Humble. Moreover, \textit{ISO/IEC/IEEE 32675} specifies the practices for operations teams, development teams and other stakeholders on collaboration for a successful building and deployment of systems \cite{32675}.
In summary, key elements include a culture of collaboration, automation of repetitive tasks, and continuous measurement. Particularly the automation of repetitive security testing and the resulting security reports motivates our research on security findings management.

Dealing with security problems in software development projects is, however, not a new research field. 
Specifically, the field of vulnerability management as part of risk management is well researched with proposed solutions ranging from enterprise- to project-level \cite{Thomas_2020, Nyanchama_2005, Wang_2009}. With concepts like unique vulnerability identifiers (CVE-IDs \cite{cve}) and vulnerability severity ratings (CVSS \cite{cvss}) this domain contributed widely employed processes and technologies that are crucial for industrial practitioners. 
For example, Farris et al. propose a framework for the management of vulnerabilities with a focus on the prioritization of vulnerability mitigation \cite{Farris_2018}. However, vulnerability management mostly focuses on problems existing in productive systems on the operations side of the software engineering lifecycle. In the best-case scenario, security problems never reach production and are instead mitigated in earlier stages, following the fast fail and shift-left mindset of DevOps. 

Towards a more holistic view, Rindell et al. propose to consider any deviation from the intended state of security as technical debt. This includes missing security features in the software and security bugs/defects alike \cite{Rindell_2019}. With the trend towards DevOps and continuous deployment, they saw the necessity for a continuous process that analyzes, validates, and keeps track of the security debt identified by tooling. An exemplary application for this use case was employed by Torkura et al. for viewing the security status history of service instances \cite{Torkura_2017}. They utilized the vulnerability management system \textit{DefectDojo} to collect security test results and visualize across multiple microservices. In contrast to other vulnerability management systems, \textit{DefectDojo} handles various types of test reports accumulated throughout the entire lifecycle. However, its main goal also represents its biggest drawback: ''The top goal of DefectDojo is to reduce the amount of time security professionals spend logging vulnerabilities'' \cite{defectdojo}. Hence, the audience is security professionals, neglecting the needs of stakeholders such as developers, or project managers to effectively collaborate on security findings management

However, DefectDojo is not the only tool supporting the management of security findings. Other representatives of this group include Faraday or Sonarqube. 
Faraday is a tool suite for the identification and management of security vulnerabilities in assets and networks \cite{Faradaysec_2023}. Their approach focuses on the analysis of existing infrastructure and the management of findings reported by these scans, indicating that solely findings from the operations stages are minded. 
In a similar fashion, Sonarqube provides the capability to conduct static code analysis and manage resulting security findings \cite{sonarqube} on their platform. The platform allows the upload of third-party reports from other tools conducting static code analysis or linting. However, the management of findings originating from external sources lacks functions like finding validation, which are exclusively enabled for internally identified security findings. Moreover, Sonarqube focuses on software code, disregarding later stages of the software development lifecycle. 

Another approach is the usage of issue trackers to manage security findings in practice \cite{Pandey_2018}. However, security findings are not equal to most other type of bugs \cite{Shahed_2011}. Even though they are mostly treated with a higher priority, they require more experienced developers to fix and are more frequently re-opened again. Moreover, the data quality of automated security testing results necessitates further investigations before they can even be considered as actual issues, minding shortcomings like False Positives. Therefore, the management of security findings has to start before any issue tracker can be utilized. Hence, we see a gap in the existing State-of-the-Art for security findings management in industrial DevOps projects.

\section{Managing Security Findings in Industry}
\label{sec:method}
The first step towards improving the security findings management process in industrial software engineering projects is the development of a comprehensive methodology supporting practitioners in their work. Following the approach of Design Science Research \cite{stol_guidelines_2020, RunesonDesign_2020}, we conducted research within Siemens AG to identify the challenges in practice. Siemens is a multinational industrial enterprise with software development activities ranging from the healthcare sector to mobility. 
In advance to the development of the methodology, we interviewed practitioners at the industry partner supporting the secure development of industrial software engineering projects. During these unstructured interviews, we asked them about issues they experienced in DevOps projects following a traditional security findings management approach. 
The results of these interviews coincide with the challenges mentioned in the introduction, logically emerging from the interconnection between software development in a domain with high-security demand and modern DevOps principles. 
These challenges and the necessity to address them are identified by commonly accepted maturity models, frameworks, and standards \cite{bsimm12, owasp_samm, safecode, sp800, 62443}. Therefore, we consider these problems as widespread and their solution as a contribution to industry and academia alike. 

Following the Design Science Research approach, a respective solution approach was designed and instantiated for each identified challenge. In the following, we present the challenges grouped by topic and their treatments considering DevOps principles as established in the methodology.

\begin{figure*}[h]
\centerline{\includegraphics[width=\linewidth]{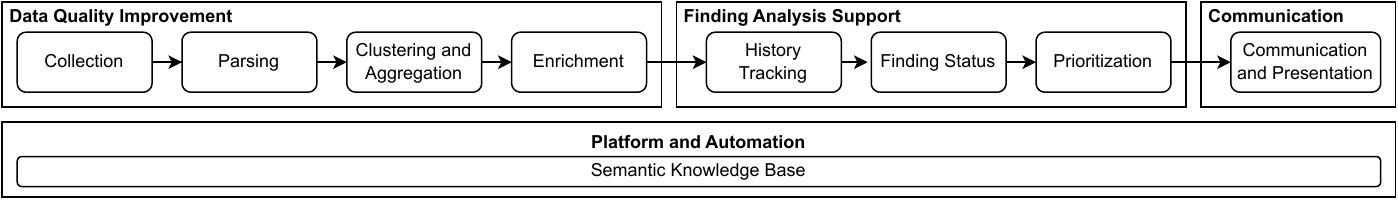}}
\caption{Methodology Design for the Security Findings Management in Industrial DevOps Projects}
\label{fig:meth}
\end{figure*}

\subsection{Data Quality Improvement}
\label{subsec:quality}
\paragraph{Challenges}
The first area of challenges we identified was related to the data quality of the source data. With various tools employing different perspectives on the product, each security report has a \textbf{different data format}. Furthermore, each security activity might be located at a different project resource resulting in \textbf{distributed source locations}. Moreover, these tools might overlap in their coverage, introducing \textbf{duplicate security findings}. Since most security tools were created by security professionals, the \textbf{terminology of findings} it generates demands practitioners to have domain expertise to understand the results. 

\paragraph{Treatments}
To improve the quality and availability of the data delivered by security activities, automated preprocessing is crucial. Since each project utilizes different tools, a tool-agnostic processing must be in place. In our methodology, the data in the form of security reports is collected from each tool and made centrally available for subsequent processing stages. Next, the data format is unified by a parsing operation utilizing a mapping between each security activity format and a common data format. 
The fields of the data model, which can be found in the supplementary material \cite{sup_material}, represent the data commonly found in security findings. To avoid duplicate security findings and provide an aggregated dataset to the later stages, our methodology follows the approach by Schneider et al. \cite{Schneider2022SemanticSC} for the clustering of security findings with semantic similarity-based techniques. This means, duplicate findings are identified by comparing textual similarities in their fields. Finally, each finding is investigated for opportunities to refine the comprised information and, therefore, simplify the comprehension of the finding's content for practitioners. This rule-based enrichment utilizes an if-then structure to identify applicable security findings that are subsequently enriched by the properties previously defined in the rule. Since the relevance of enrichment is highly project-specific, almost no general rules are given. The only default rule in our methodology is a tool-dependent explanation of how each finding was identified (E.g., \textit{The tool $<$Tool$>$ identified this finding, by querying each of your dependencies in public vulnerability databases. If a vulnerability is found for a component, it creates a finding.}). 
The solution approach for the \textit{Data Quality Improvement} can be found on the left side of Figure~\ref{fig:meth}. 

This stage can be fully automated during the project. Solely the parser mapping each source activity and rules for enrichment must be defined upfront. 

\subsection{Finding Analysis Support}
\label{subsec:analysissupp}

\paragraph{Challenges}
To enable practitioners during their work on security findings, respective insights into the current state of security must be accumulated. One indicator of the relevance and prevalence of security findings is the identification history. This comprises the occurrence of findings in security reports, including factors like first or last identification. Moreover, each security finding reaches different states during its lifecycle that must be maintained consistent throughout continuously new data being reported by security activities. 
Finally, the treatment of each finding depends on multiple factors, including but not exclusively being the severity of security findings. Since the time available for changes to the software is limited by the size and maturity of the development team, prioritizing the response to findings is crucial. 

\paragraph{Treatments}
To support practitioners in analyzing and responding to security findings, three activities are aided by the methodology. First, the methodology tracks the history of each finding, covering all occurrences in security activity reports. This provides historical data on each finding, including the first and last identification, its frequency, and distribution across sources. 
Moreover, the status of each finding is tracked. This status represents the result of a finding verification and persists throughout incremental finding changes. The available values for the finding status comprise: ''Open'', ''In Work'', ''False Positive'', ''Invalid'', ''Accepted'', ''Solved'', ''On Hold'' and ''Disappeared''. A finding's status persists either permanently or under certain circumstances (e.g. finding was not found in the last two reports). By default, every finding receives the status ''Open'', which is solely changed to ''Disappeared'' if the finding was not found in the last report. ''Disappeared'' is furthermore the only status that cannot be assigned by users of the methodology. 
Finally, each finding must be prioritized to cope with reduced available implementation time per iteration. Our methodology supports this prioritization process by providing a common finding severity score and a project-dependent prioritization score for each finding. We follow the approach by Voggenreiter and Schöpp \cite{Voggenreiter2023Prio} for the calculation of both scores. This implies that the severity of each finding is calculated by applying activity-based models to the findings data, which is afterward manually refined by the project team or stakeholders to present the importance of caring about this finding on a numeric scale. 
Finally, our methodology requires that all data resulting from the above-mentioned process steps is documented and traceable at any time.
The solution approach for the \textit{Finding Analysis Support} can be found in the middle of Figure~\ref{fig:meth}. 

\subsection{Security Feedback Communication}
\label{subsec:communication}
\paragraph{Challenges}
In order for practitioners to fulfill their tasks based on the security findings, knowledge about findings must be communicated to them.
Since the subsequent actions depend on the role of the knowledge-receiving entity, the \textbf{communication must be tailored towards the user's role}. Minding the importance of cross-domain collaboration, all \textbf{stakeholders and team members require access to the security findings data}. However, the data must also be \textbf{accessible to automated processes}, e.g., to provide cross-project correlations. Finally, the \textbf{data must be continuously available}, so that the team can work on the management of security findings whenever time allows it.  

\paragraph{Treatments}
To communicate the security feedback, our methodology defines a common interface to practitioners and processes. 
For the processes, an automated interface in the form of an API is necessary, while human users require a visual approach using a webinterface. Both interfaces provide access to all security findings and their related information. Moreover, role-based views on the data provide users with the information necessary for their tasks. To contribute to a tailored communication strategy, a specific view for developers focusing on the solution aspects of the management is necessary. 

This stage can be fully automated, except for additional role-based views that might be necessary. The solution approach for the \textit{Security Feedback Communication} is depicted on the right side of Figure~\ref{fig:meth}. 

\subsection{Platform and Automation}
\label{subsec:automate}
\paragraph{Challenges}
With all previous solution approaches defined, performing them manually is unfeasible. Hence, all \textbf{steps must be orchestrated and automated} as far as possible. The information acquired during the methodology must be \textbf{centrally documented and available}, to access the information at any time. Moreover, our methodology solely presents a snapshot of generally existing challenges. Further process steps might be employed at more sophisticated projects, therefore requiring the platform to be \textbf{customizable} to support these as well. 

\paragraph{Treatments}
Finally, all preceding solution approaches must be orchestrated and automated within a common platform to comply with DevOps principles and avoid manual, repetitive work. For this platform, we follow the idea of using a semantic knowledge base as a platform for the management of security findings \cite{Voggenreiter2022UsingAS}. This knowledge base represents a data source for each project which maintains consistent information throughout new reports being added by sources or data being changed by external influences like developers. We extend their work by adding the concept of queries to the knowledge base, introducing temporarily computed views on the data that are customized according to time and necessary information. This results in the following four concepts:
\begin{itemize}
    \item Belief: Any type of data ranging from security reports to the prioritization score of single findings
    \item Rules: Logic, describing how to derive new Belief from existing Belief
    \item Query: Computational Rules, which provide information based on the current elements of Belief in the knowledge base
    \item Data Storage: Underlying storage element, storing Belief, Rules, and Queries
\end{itemize}
Applied to our previous solution approaches, each piece of information is considered as a belief. Every data processing step is a rule transforming, e.g., a security report to multiple security findings by applying parsing rules. 

\begin{figure}[h]
\centerline{\includegraphics[width=\linewidth]{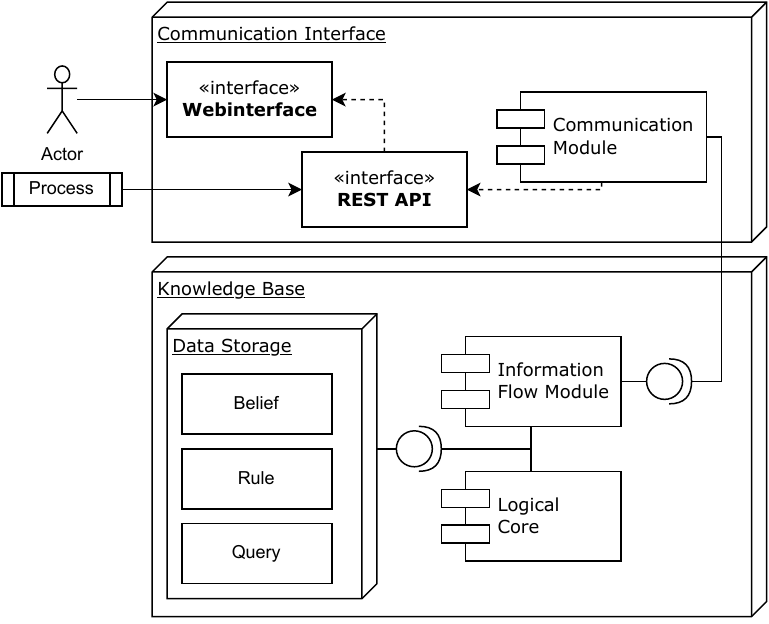}}
\caption{Components of the \textit{Security Flama}}
\label{fig:implement}
\end{figure}
Implementation-wise, the knowledge base uses Elasticsearch as \textit{Data Storage}. \textit{Belief} is stored as documents in the search engine while \textit{Rules} are coded in Python. The \textit{Queries} are Python snippets, including Elastic Query DSL requests to access belief in the knowledge base. The knowledge base is maintained consistent by a logical core that ensures that new belief is derived from existing belief and falsified data is corrected after new insights occur. 
Access to the data is provided by two interfaces. Automated processes can access the knowledge base via a RESTful API, while users can access a visual representation via a website. The website includes a customized, role-based dashboard, a sort- and filterable list of all findings, and a separate page for each finding. Examples for these pages can be found in the supplementary material \cite{sup_material}. We call this instantiation of the methodology the \textit{Security Flama}. Its components are depicted in Figure~\ref{fig:implement}. 

\subsection{Processing Example}
\label{subsec:methexample}
A typical workflow for the \textit{Security Flama} starts with new security reports being generated in a CI/CD pipeline with automated security tests. Using a separate job in the pipeline, the resulting reports are uploaded to the REST API and forwarded to the Information Flow Module of the \textit{Security Flama} (see Figure~\ref{fig:implement}). This module adds the new reports as instances of the belief class ''Security Report'' to the data storage and informs the logical core about a change in the dataset. The logical core checks, whether any rule is triggered by a new instance of this class and identifies the rule, which parses security reports to distinct findings. Executing the Python code of this rule, the logical core uses the new security report as input and returns new instances of the belief class ''Security Finding'' representing the parsed, distinct findings. The newly constructed security findings are added to the data storage again and the logical core re-evaluates whether any rule is triggered by new instances of security findings in the data storage. This process is iteratively executed, with the logical core analyzing whether any rule can be used to derive new information and the execution of applicable rules. In our example, this results in the subsequent execution of the deduplication and aggregation using Latent Semantic Indexing, rule-based enrichment, history and status tracking, and the modeling of severity and priority for each security finding. This results in additional instances of belief classes being added to the knowledge base. Afterward, the knowledge base reaches a consistent state, as no additional belief can be derived from the newly added security reports. This implies that the processing of the uploaded security reports concluded. 

After the automated tests have been conducted, project team members might be interested in the currently existing security findings. Therefore, the next part of the workflow would be a user of the \textit{Security Flama} accessing the webinterface, which again forwards the request to the Information Flow Module. This module computes the response to the request by accessing the knowledge base queries, which again query the belief instances, constructing answers to questions like ''What are the ten findings with the highest priority'' or ''How many findings are currently open''. This knowledge is presented by the webinterface based on the response of the Information Flow Module. Any potential user input, e.g., refinement on the priority of security findings, is also implemented via the communication interface and stored as belief. 
Finally, users would enter any high-priority security finding into their project-specific issue tracker or backlog to track its implementation. 
\section{Evaluation}
\label{sec:evaluation}
Using the methodology described in the last section, we collaborated with our industry partner once again to verify our methodology as treatment to the challenges in practice. In this section, we describe this evaluation by using the \textit{Security Flama} as an instance of it.

\subsection{Evaluation Planning}
\label{subsec:planning}
The close relation of our methodology to the effectiveness of the software development lifecycle necessitates an evaluation covering quantitative performance indicators of a real-world project as well as qualitative perception data by users of the methodology. Consequently, we evaluate the \textit{Security Flama} through the following research questions (RQ):
\begin{itemize}
\item\textbf{RQ1: How does the usage of the \textit{Security Flama} impact the development process indicators?}
\item \textbf{RQ2: Is the \textit{Security Flama} perceived as useful by the project team for managing security findings?}
\item \textbf{RQ3: What are the benefits and limitations of applying the \textit{Security Flama} for project stakeholders?}
\end{itemize}

Towards this goal, we conduct a case study, following the State-of-the-Art for empirical software engineering and case study research \cite{Runeson_2009, mendez_2018, Santiago_2017, Seaman2008} minding challenges and domain experience \cite{othmane_jaatun_weippl_2017}. 

As subjects, we require ongoing software development projects at our industry partner. Each project has to follow DevOps principles, run for more than one year, utilize automated security tests with a subsequent security findings management process, and contain weekly code changes. 
Our partner provided us with two independent projects in the R\&D domain fulfilling our demands, consisting of two (Project A) and three developers (Project B), respectively. 
Both projects employed a Kanban board for task planning and the issue tracker integrated into the code repository to manage coding-related problems. Their existing security findings management process comprises the manual review of security testing reports in the CI environment and a subsequent discussion of these results in the weekly project team meeting, potentially resulting in a new issue in their backlog. 
Since the management of security findings is up to each project, this strategy is not representative of all projects at our partner. Projects utilize a variety of combinations between different code repositories, issue trackers, and task planners. 

\begin{table}
\caption{Quantitative Data Collected}
\begin{center}
\begin{tabular}{|c|}
\hline
  \textbf{Indicator}\\
\hline
Number of security activities producing findings in the project.\\ 
 \hline
Number of security tools providing data to the \textit{Security Flama}. \\ 
 \hline
Number of reports created per week. \\
 \hline
Number of raw findings identified per week. \\
 \hline
Number of aggregated findings per week. \\
 \hline
Number of new findings per week. \\
 \hline
Number of findings with status X (Open, False Positive, ...). \\
 \hline
Number of findings with severity X (Critical, High, ...) \\
 \hline
Number of user input per week (Status, Prioritization). \\
 \hline
\end{tabular}
\label{tab:quant}
\end{center}
\end{table}

Each project follows a rigorous protocol during the evaluation:
\begin{enumerate}
    \item Integration of the \textit{Security Flama}
    \item Passive, continuous collection of statistics from the project 
    \item Introduction of the \textit{Security Flama} to the project team
    \item Data collection with five iterations of questionnaires
    \item Conclusion of data collection with interview
    \item Ramp-Down
\end{enumerate}
Initially, the \textit{Security Flama} is integrated into the ongoing projects and collects security reports resulting from security tests of the main branch with no introduction of the methodology to the project teams. In the first four weeks after integration, the projects are passively monitored without intervention to achieve a baseline of data. This monitoring is implemented by a weekly collection of the quantitative data listed in Table~\ref{tab:quant}. 
Afterward, the methodology is presented to the project team, including developers and other stakeholders, and access to the user interface is established for all users. Furthermore, the case study and its goals are presented to the subjects.  
To capture potential changes in the perception of the \textit{Security Flama} over time \cite{Wagner_2020}, each team member is interviewed bi-weekly, focusing on the experience of the past two weeks. The interview consists of a questionnaire comprised of open and binary questions as well as questions on a Likert Scale defined from 1 - ''Strong Disagree'' to 6 - ''Strong Agree'', avoiding any neutral answer. 
After five iterations of the reoccurring interviews, the evaluation is concluded by a final interview. This interview assesses the subjective \textit{usability}, \textit{benefits}, and \textit{limitations} of the \textit{Security Flama} utilizing a distinct questionnaire. 
The initial presentation, both questionnaires and the interview guide can be found in the supplementary material \cite{sup_material}. 

\subsection{Evaluation Results}
\label{subsec:results}
The results presented in this section are published in accordance with the collaboration agreement with our industry partner. Consequently, the results are partially anonymous to comply with this agreement. 

\noindent\textbf{Quantitative Results:}\\
\noindent Over the course of the evaluation, we collected the quantitative evidence on 16 occasions. 
In both projects all security activities were automated, implying that the amount of security activities equals the amount of security tools. In Project A, a secret scanning tool, a static code analysis, and a third-party vulnerability testing tool produced the security reports managed by the \textit{Security Flama}. In Project B, an additional, tailored third-party vulnerability testing tool was added on February 15 during the evaluation period, after a gap in the testing coverage was identified during the review of security findings. 

\begin{figure}[h]
\centerline{\includegraphics[width=\linewidth]{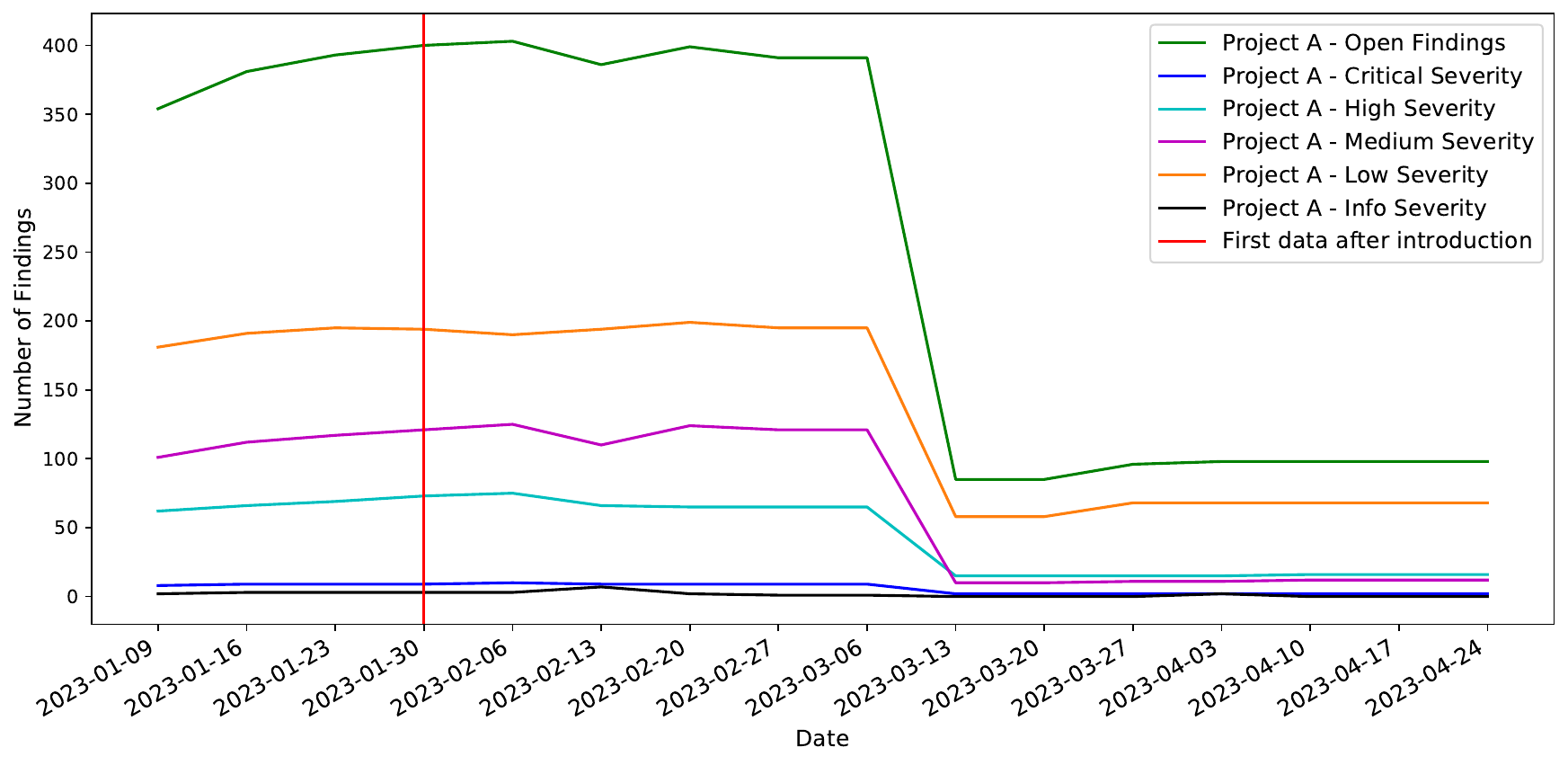}}
\caption{Severity of Findings in Project A}
\label{fig:sev_A}
\end{figure}

\begin{figure}[h]
\centerline{\includegraphics[width=\linewidth]{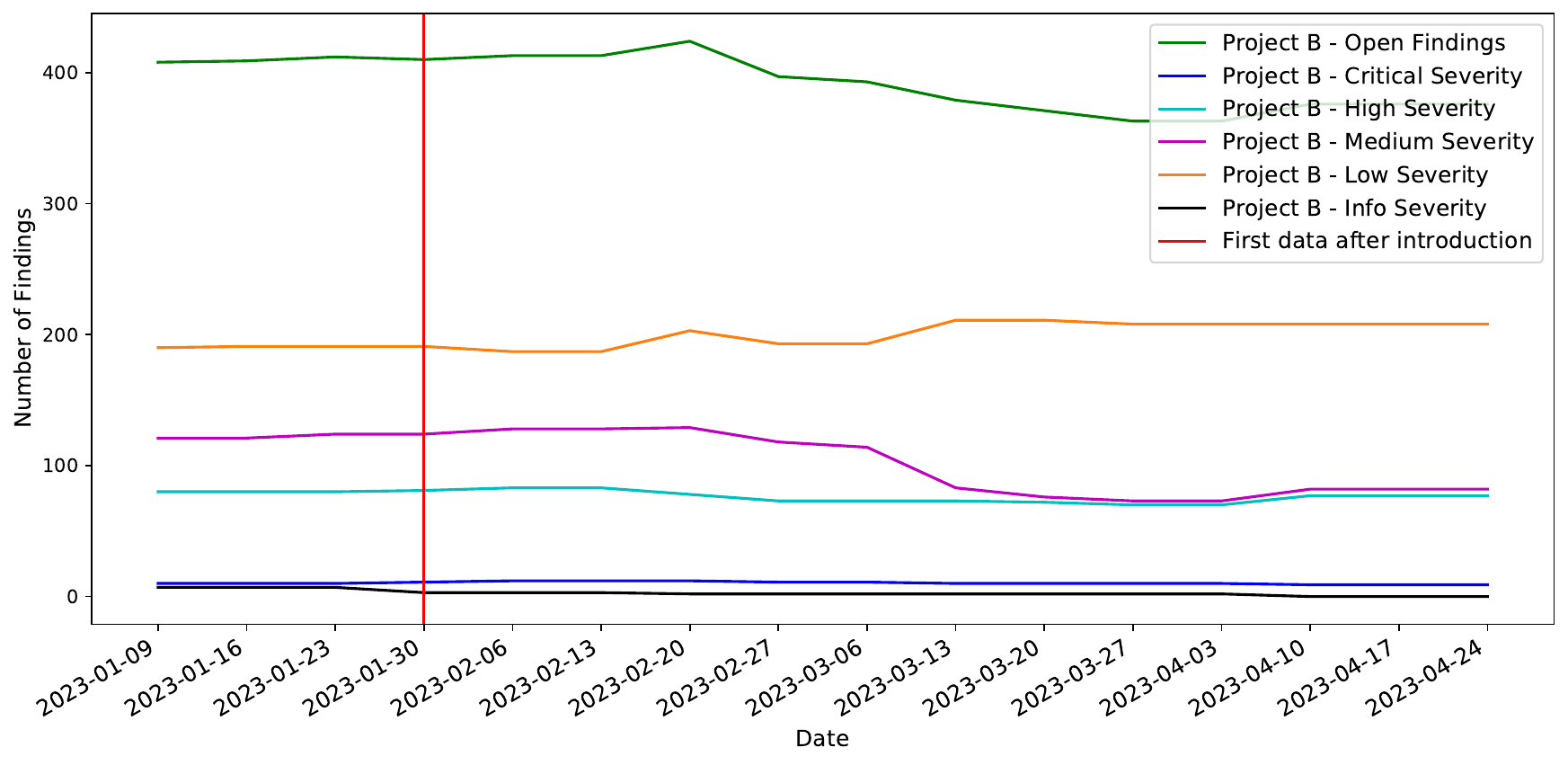}}
\caption{Severity of Findings in Project B}
\label{fig:sev_B}
\end{figure}

\begin{figure}[h]
\centerline{\includegraphics[width=\linewidth]{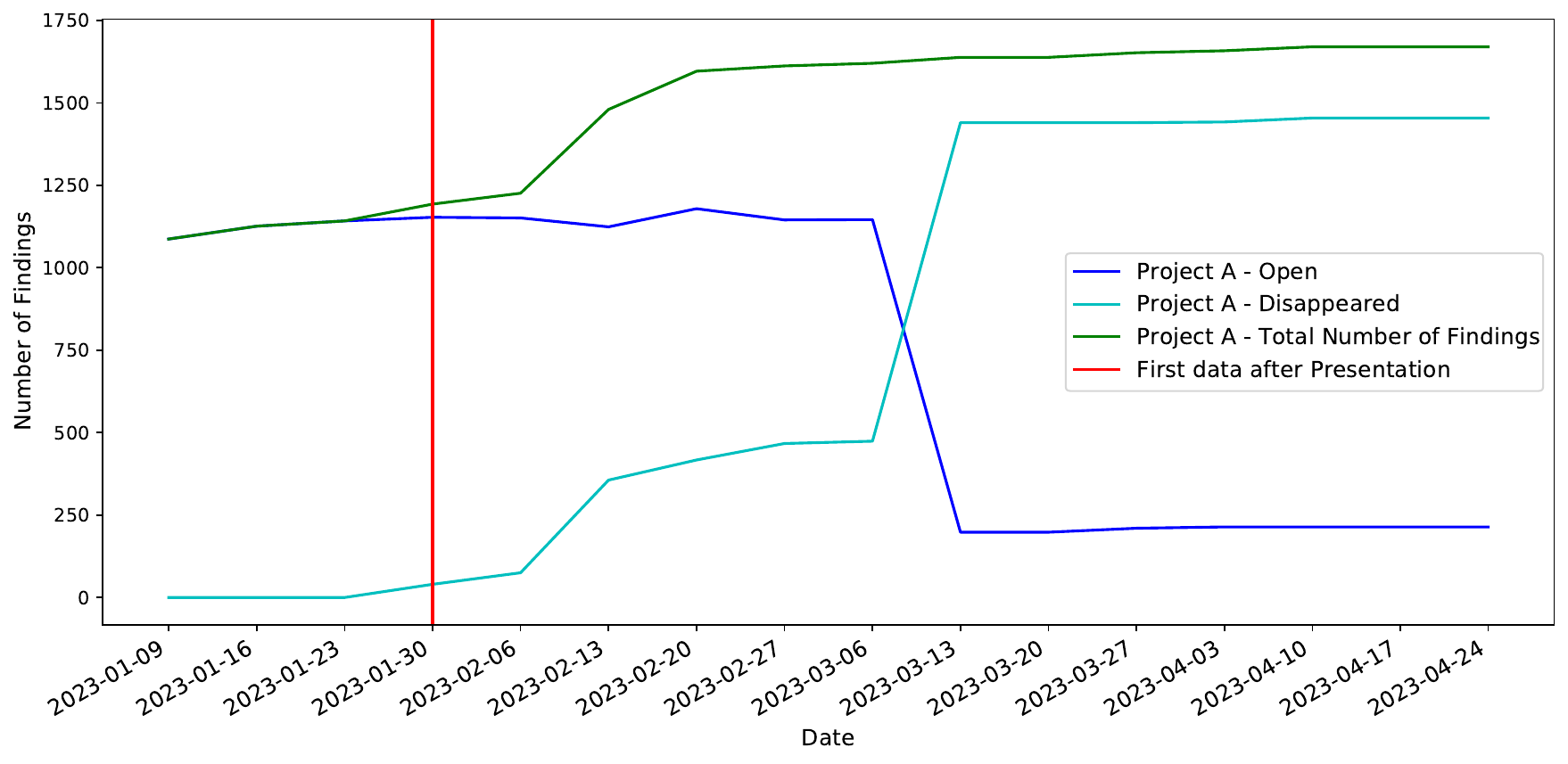}}
\caption{Findings Status of Project A}
\label{fig:quant_A}
\end{figure}

\begin{figure}[h]
\centerline{\includegraphics[width=\linewidth]{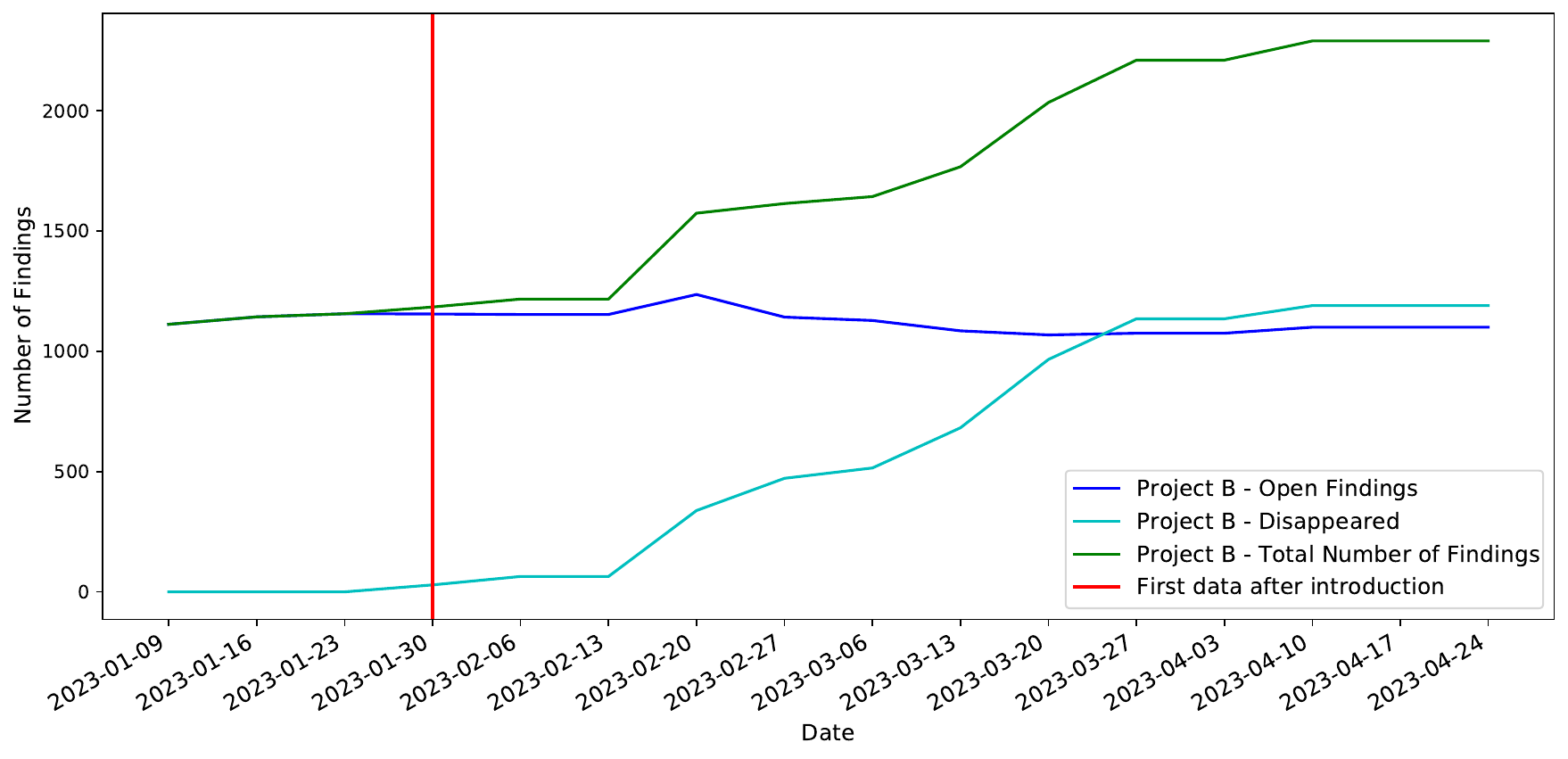}}
\caption{Findings Status of Project B}
\label{fig:quant_B}
\end{figure}

\begin{figure}[h]
\centerline{\includegraphics[width=\linewidth]{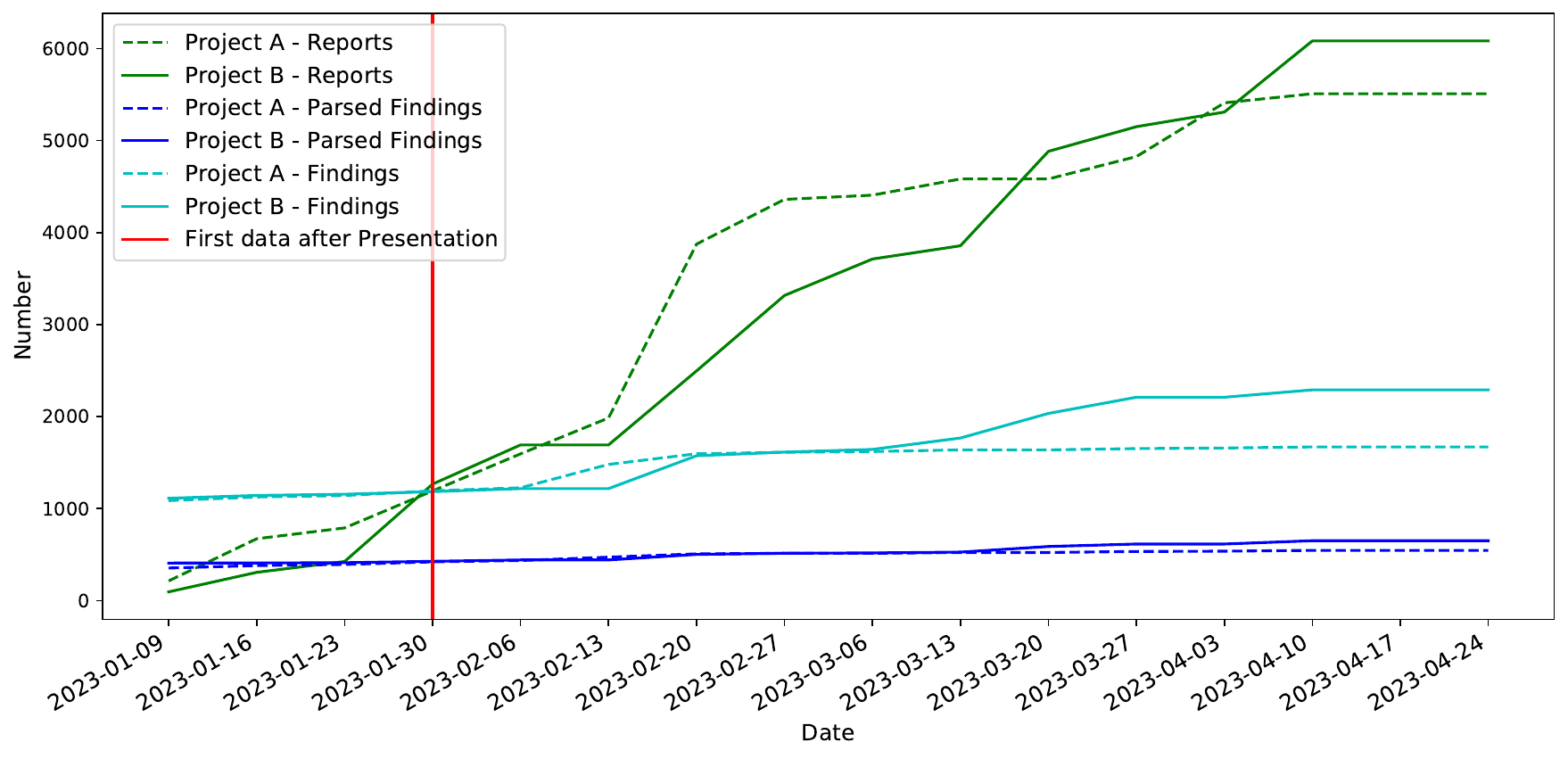}}
\caption{Finding Statistics of both Projects}
\label{fig:quant_overall}
\end{figure}

{\footnotesize
\rowcolors{2}{gray!25}{white}
\begin{table*}
\setlength{\tabcolsep}{2pt}
\caption{Quantitative Data of Projects A and B}
\label{tab:data_comb}
\begin{center}
\begin{tabular}{ *{31}{|c} | }
\rowcolor{gray!50}
     date &  \multicolumn{2}{p{11mm}|}{number reports} &  \multicolumn{2}{p{11mm}|}{number findings} &  \multicolumn{2}{p{9mm}|}{parsed findings}  &  \multicolumn{2}{p{9mm}|}{new findings}  &  \multicolumn{2}{p{10mm}|}{open findings} &  \multicolumn{2}{p{8mm}|}{number prio} &  \multicolumn{2}{p{7mm}|}{sever. crit} & \multicolumn{2}{p{7mm}|}{sever. high} &  \multicolumn{2}{p{9mm}|}{sever. medium} & \multicolumn{2}{p{9mm}|}{sever. low} &  \multicolumn{2}{p{6mm}|}{sever. info} &  \multicolumn{2}{p{11mm}|}{status Open} &  \multicolumn{2}{p{11mm}|}{status Disappeared} &  \multicolumn{2}{p{6mm}|}{status FP} &  \multicolumn{2}{p{7mm}|}{status Accep.}\\
2023-01-09 &             214&95&        1087&1112&      354&408&        354&408&        354&408&        0&0&        8&10&       62&80&      101&121&        181&190&        2&7&        1087&1112&      0&0&        0&0&        0&0 \\
2023-01-16 &             672&307&       1126&1143&      381&409&        27&1&           381&409&        0&0&        9&10&       66&80&      112&121&        191&191&        3&7&        1126&1143&      0&0&        0&0&        0&0 \\
2023-01-23 &             790&425&       1142&1156&      393&412&        12&3&           393&412&        0&0&        9&10&       69&80&      117&124&        195&191&        3&7&        1142&1156&      0&0&        0&0&        0&0 \\
2023-01-30 &             1192&1266&     1193&1184&      421&427&        28&15&          400&410&        0&0&        9&11&       73&81&      121&124&        194&191&        3&3&        1153&1155&      40&29&      0&0&        0&0 \\
2023-02-06 &            1595&1692&      1226&1217&      436&442&        15&15&          403&413&        0&0&        10&12&      75&83&      125&128&        190&187&        3&3&        1151&1153&      75&64&      0&0&        0&0 \\
2023-02-13 &            1985&1692&      1480&1217&      472&442&        36&0&           386&413&        0&0&        9&12&       66&83&      110&128&        194&187&        7&3&        1124&1153&      356&64&     0&0&        0&0 \\
2023-02-20 &            3874&2496&      1596&1574&      508&502&        36&60&          399&424&        0&0&        9&12&       65&78&      124&129&        199&203&        2&2&        1179&1236&      417&338&    0&0&        0&0 \\
2023-02-27 &            4360&3316&      1612&1614&      514&515&        6&13&           391&397&        0&1&        9&11&       65&73&      121&118&        195&193&        1&2&        1145&1142&      467&472&    0&0&        0&0 \\
2023-03-06 &            4406&3712&      1620&1643&      514&518&        0&3&            391&393&        0&0&        9&11&       65&73&      121&114&        195&193&        1&2&        1146&1128&      474&515&    0&0&        0&0 \\
2023-03-13 &            4582&3856&      1638&1767&      523&527&        9&9&            85&379&         0&0&        2&10&       15&73&      10&83&          58&211&         0&2&        198&1085&       1440&682&   0&0&        0&0 \\
2023-03-20 &            4582&4882&      1638&2034&      523&588&        0&61&           85&371&         0&0&        2&10&       15&72&      10&76&          58&211&         0&2&        198&1068&       1440&966&   0&0&        0&0 \\
2023-03-27 &            4825&5150&      1652&2210&      534&615&        11&27&          96&363&         0&1&        2&10&       15&70&      11&73&          68&208&         0&2&        210&1075&       1440&1135&  1&0&        1&0 \\
2023-04-03 &            5409&5309&      1658&2210&      537&615&        3&0&            98&363&         0&1&        2&10&       15&70&      11&73&          68&208&         2&2&        214&1075&       1442&1135&  1&0&        1&0 \\
2023-04-10 &            5508&6082&      1670&2290&      546&651&        9&36&           98&376&         0&1&        2&9&        16&77&      12&82&          68&208&         0&0&        214&1100&       1454&1190&  1&0&        1&0 \\
2023-04-17 &            5508&6082&      1670&2290&      546&651&        0&0&            98&376&         0&1&        2&9&        16&77&      12&82&          68&208&         0&0&        214&1100&       1454&1190&  1&0&        1&0 \\
2023-04-24 &            5508&6082&      1670&2290&      546&651&        0&0&            98&376&         0&1&        2&9&        16&77&      12&82&          68&208&         0&0&        214&1100&       1454&1190&  1&0&        1&0 \\
\end{tabular}
\end{center}
\end{table*}
}
The aggregated data collected in both projects can be found in Table~\ref{tab:data_comb}. The results of Project A are depicted on the left of each column, while the results of Project B are shown on the right. 
\begin{table}
\caption{Quantitative Statistics per Week}
\begin{center}
\begin{tabular}{|l|c|c|c|c|c|c|}
\hline
\textbf{Indicator} &  \multicolumn{3}{c|}{\textbf{Project A}} & \multicolumn{3}{c|}{\textbf{Project B}} \\
\hline
& \textbf{Min} &  \textbf{Max} &  \textbf{Avg} & \textbf{Min} &  \textbf{Max} &  \textbf{Avg}\\
\hline
\#Security Activities & 3 & 3 & 3 & 3 & 4 & 4\\ 
 \hline
\#Security Tools & 3 & 3 & 3 & 3 & 4 & 4\\ 
 \hline
\#Reports& 0 & 1889 & 344 & 0 & 1026 & 380 \\
\hline
\#Raw Findings& 0 & 1087 & 104 & 0 & 1112 & 143 \\
\hline
\#Aggregated Findings& 0 & 354 & 34 & 0 & 408 & 40 \\
\hline
\#New Findings& 0 & 36 & 12 & 0 & 61 & 17 \\
\hline
\#User Input& 0 & 2 & 0.25 & 0 & 1 & 0.125\\
\hline
\end{tabular}
\label{tab:quant_res}
\end{center}
\end{table}
The statistics calculated about the analyzed indicators can be found in Table~\ref{tab:quant_res}. We consider a ''New Finding'' to be first identified in the last seven days, hence since the last data point. We excluded the first date from this calculation, as initially all findings are new. All values are rounded to the nearest lower integer, except for the ''User Input per Week''. 
The ''Amount of findings with severity X / status X'' is excluded from the table due to its verbosity. Instead, the severity of findings over time are presented in Figure~\ref{fig:sev_A} for Project A and in Figure~\ref{fig:sev_B} for Project B. 
Automated changes towards the finding status are presented in Figure~\ref{fig:quant_A} for Project A and in Figure~\ref{fig:quant_B} for Project B. 
The only manual change in finding status appeared in Project A in the week of March 20. Two findings were assigned the status ''False Positive'' and ''Accepted''.  
Moreover, the number of reports, raw findings, and aggregated findings of both projects can be found in Figure~\ref{fig:quant_overall}. 

\begin{figure}[h]
\centerline{\includegraphics[width=\linewidth]{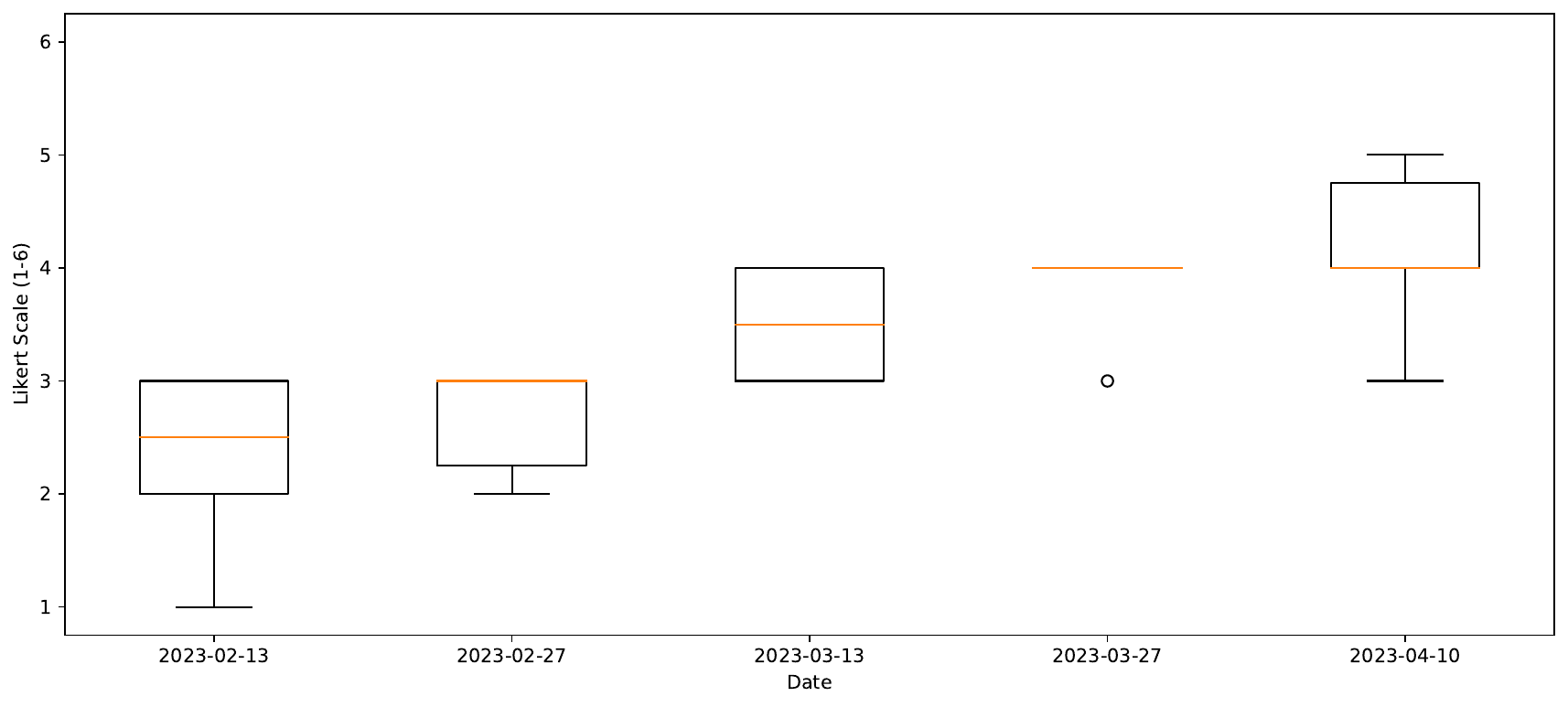}}
\caption{Perception on the Transparency of the \textit{Security Flama} over Time}
\label{fig:trans}
\end{figure}

\noindent\textbf{Qualitative Results - Repeated Interviews:}\\
\noindent In addition to the quantitative data, the reoccurring interviews with the project teams supplied our evaluation with qualitative information about the performance of the \textit{Security Flama}. In summary, each team member was interviewed five times based on the reoccurring questionnaire and one time with the final questionnaire between February 13 and April 28. One developer left Project B in March, resulting in the completion of just two of five interviews for this subject. Especially during the first interview sessions, the subjects were hesitant to answer questions, as they felt unable to do so with the limited time they spent on using the \textit{Security Flama}.   
Most answers on a Likert scale were answered positively with a ''5-Agree'' or ''6-Strong~Agree''. Only questions 17 and 12 were exceptions. Question 17, asking whether the subject was restricted in its work by the \textit{Security Flama} in the last two weeks, was answered with ''1-Strong~Disagree'' consistently. 
Another exception was question 12, asking about the transparency of how the \textit{Security Flama} works. This question showed a development over time, depicted in Figure~\ref{fig:trans}. The answers to this question ranged from ''1-Strong~Disagree'' to ''5-Agree''. 

Since answers to the open questions contained confidential data on active vulnerabilities, these are abstracted for presentation. The subjects mentioned the overview of the currently existing security findings in the product as notably useful and the main benefit of the \textit{Security Flama}. The referenced aspects ranged from directly identifying new findings, over having an aggregated view on the findings list to the common severity scale allowing to compare findings with each other. However, all subjects mentioned these aspects in the context of gaining ''overview''. According to the interview results, the \textit{Security Flama} itself was mainly used in preparation for or during team meetings to align the response to security findings in the team. The only limitations mentioned during the interviews were bugs in the implementation of the webinterface. Similarly, the \textit{Security Flama} improvements and general comments of subjects exclusively addressed potential changes in the webinterface implementation.  

\noindent\textbf{Qualitative Results - Final Interviews:}\\
\noindent To conclude the evaluation, six final interviews were conducted with the remaining team members. The results of these interviews are separated into the three assessed properties of the \textit{Security Flama}. Its \textit{usability}, the \textit{benefits} of using it, and its \textit{limitations} in the project.
One subject felt uncomfortable answering the interview questions due to their insufficient direct interaction with the \textit{Security Flama}, resulting in five interviews. 

All subjects confirmed the usability of the \textit{Security Flama}, describing it as ''highly beneficial'' or ''extremely useful''. The collectively mentioned reason is the ''up-to-date insight on the security state'' of a product, covering ''newly introduced findings'' and allowing for ''insights otherwise impossible''. 
The second most mentioned factor in usability is the simplicity of accessing the security findings data. 

As benefits of the \textit{Security Flama} our subjects mentioned various aspects. 
The most commonly mentioned one, was its flexibility to tailor the tool to the needs of the project in terms of security activities used and user input. The evidence-based overview on the security state of the software was again mentioned by all subjects as a benefit. Moreover, some components of the \textit{Security Flama} were mentioned as beneficial, including the common datamodel, the aggregation of findings, the calculation of severity on a common scale, and the ''role-tailored communication of data''. 

Limitations of the \textit{Security Flama} mentioned by the subjects can be categorized into three clusters. 
The first cluster addresses the limitations imposed on the project when using the \textit{Security Flama}. The subjects described the effort to actively manage findings as ''exhaustive'' and often forgot about it. One subject wished for a more proactive communication approach like email notifications if critical findings occur, which the team could subsequently react to. Moreover, this team also managed every finding that should be fixed in the backlog. Thus, findings were prioritized against other backlog items, resulting in the subject's perception of ''duplicate housekeeping''. Consequently, the manual effort to interact with the \textit{Security Flama} was seen as a limitation for the project. 
The second cluster addressed shortcomings of the \textit{Security Flama} and potential procedural improvements. The subjects mentioned the poor visibility of how the data is processed by the \textit{Security Flama} and the complexity of adding priority scores with numbers as shortcomings. Moreover, they perceived the lack of a long-term success overview for all roles to be detrimental to the team's motivation. One subject observed the necessity for a change in the finding status system. According to the subject, the finding status was not adapted in their project, as team members wanted to confirm a finding fix by checking whether it disappeared. 
Finally, multiple subjects gave recommendations on how to improve the implementation of the webinterface. However, these were not related to the \textit{Security Flama} directly, but just related to aspects of the communication. 
\section{Discussion}
\label{sec:discuss}
In this section, we discuss the results of our evaluation, starting with general discoveries and focusing on answering the research questions in a second step. Finally, we address threats to the validity of our research. 

\subsection{General Analysis of Evaluation Results}
\label{subsec:genanalysis}
The data acquired during the evaluation shows multiple features, that must be discussed upfront before addressing the research questions.

First, we have to correct our initial assumption that all team members and stakeholders interact with the methodology directly. Contrary to our initial assumption, one subject accessed the data exclusively through other members during team meetings or on specific occasions. These subjects, either missing the time or interest to work with security findings, represent a new type of methodology user that has not been considered yet.

To interpret the finding status, it is important to note that each location of a finding in the product has a distinct status, and numerous locations can be aggregated within the same finding. Consequently, the sum of all status values is equal to the number of raw findings presented in ''number findings''. Since each aggregated finding is assigned a severity, the sum of all severity levels is equal to the number of aggregated findings presented in ''parsed findings''.

Moreover, an increase in security findings can be observed for Project B on April 20 (see Figure~\ref{fig:quant_B}). This is explained by the additional security tool being added on February 15. 

As seen in Table~\ref{tab:data_comb}, the data on the ''2023-04-10'', ''2023-04-17'', and ''2023-04-24'' shows the exact same values. This is due to the public Easter Holidays, resulting in no code changes within both projects. These missing data changes can also be observed, whenever code changes and tests only happened on project branches since only the results affecting the main branch were used for findings management. 

Finally, it is crucial to mention that the number of reports, raw and aggregated findings can only increase over time as depicted in Figure~\ref{fig:quant_overall}. This is due to the methodology's history tracking, which persistently stores all external data and the belief derived from it.  

\subsection{Answers to Research Questions}
\label{subsec:analysis}
Given the data presented in the last section, we answer the research questions in this subsection. 

\noindent\textbf{RQ1: How does the usage of the \textit{Security Flama}
impact the development process indicators?}
The impact of the \textit{Security Flama} can be visually recognized in Figure~\ref{fig:quant_A} and Figure~\ref{fig:quant_B}. First, we observe that the aggregation diminishes the number of findings by at least two-thirds in both projects (see Figure~\ref{fig:quant_overall}). Moreover, in the week of March 6, one project team responded to multiple security findings based on the insights provided by the \textit{Security Flama}, which drastically reduced the number of findings in the project. According to the interviews, this was possible due to the unique insights provided by the \textit{Security Flama}. 

While both projects show a reduction of open, unsolved findings over the course of the evaluation (green line in Figures~\ref{fig:sev_A},~\ref{fig:sev_B}), this decrease can only be considered as substantial in Project A. Therefore, we conclude that the success of the \textit{Security Flama}, as any security tool,  highly depends on its acceptance and the importance of software security to the stakeholders. 
Also, the distribution of finding severity shown in Figure~\ref{fig:sev_A} and Figure~\ref{fig:sev_B} within both projects did not change notably. 

\begin{mdframed}[backgroundcolor=lightgray]
One project showed a substantial improvement in the process indicators when using the \textit{Security Flama}. The impact on the other project was marginal due to the lack of adoption. 
\end{mdframed}

\noindent\textbf{RQ2: Is the \textit{Security Flama} perceived as useful by the project team for managing security findings?}
During the interviews, all subjects described the \textit{Security Flama} as useful for the project and their work. Especially in Project A, the subjects argued that the \textit{Security Flama} enabled them to identify quick wins for the reduction of the overall number of findings (see Figure~\ref{fig:quant_A}). This considerable impact on the security of the software was traced to evidence-based visibility provided by the \textit{Security Flama} in the reoccurring and final interviews. 

Even though the subjects claim that the \textit{Security Flama} was perceived as useful, we see some restrictions when analyzing the quantitative data. The overall user input provided to the \textit{Security Flama} is remarkably low. As seen in Table~\ref{tab:data_comb}, only one finding was assigned a priority score and two findings received a manually assigned status. The interview data gives various explanations for this behavior. Our subjects explain the absence of priority data with (1) the high complexity of using numeric scores as user input and (2) the approach of both projects in which only high and critical findings can and will be addressed at their current development stage. Consequently, a separate, more granular prioritization of findings was unnecessary in both projects. The low assignment of a finding status, on the other hand, was blamed on missing comfort functions in the web interface and the need for the automated confirmation of a finding not being reported again. Due to the binary decision about which findings are even considered relevant, one project wished for a mass tagging of all findings with severity medium and beneath as ''accepted'' by default. This would have affected 325 aggregated findings and therefore added more than 1000 status changes to the dataset of Project A at that time (2023-02-20).  

We conclude that even though the \textit{Security Flama} was perceived as highly useful, this perception seems to rely uniquely on the aspects that were also utilized. We assume that the prioritization would have been not perceived as useful if its usage was forced as part of the process. However, the ability to decide which aspects of the methodology are relevant for the project at its current state ensured its usefulness to the projects. 

\begin{mdframed}[backgroundcolor=lightgray]
Even though the \textit{Security Flama} is perceived as highly useful, we believe that this perception relies on the ability to avoid interaction with inconvenient aspects of the methodology like manual prioritization.
\end{mdframed}

\noindent\textbf{RQ3: What are the benefits and limitations of applying the \textit{Security Flama} for project stakeholders?}
The benefits and limitations of the \textit{Security Flama} have been shown to be the most insightful part of the interviews. 

Most subjects mentioned bugs in the web interface as limitations or shortcomings of the \textit{Security Flama}. This shows the importance of a high-quality communication strategy, as it represents the single point of contact for practitioners to interact with the \textit{Security Flama}. 
The communication strategy we followed of using a single web interface was described as insufficient during the final interview. One subject wished for a complementary proactive strategy with emails being sent, whenever new or critical findings arise as reminders to access the web interface. Moreover, integration with the product backlog was missing according to the subjects, resulting in duplicate housekeeping of security findings. Due to the broad variety of backlogs and issue trackers existing in practice, this was not implemented in the initial version of the \textit{Security Flama}. Consequently, the implementation of our communication strategy needs to be further improved to better support the needs of practitioners.  

However, not every limitation focused on the communication strategy. The current approach for prioritizing security findings was considered as not relevant for one project as prioritization happens in the respective backlog against other non-functional and functional requirements. Subjects of the other project perceived the user input necessary for prioritization as too complex. Even though we acknowledge the necessity of reducing the input complexity, we still believe that prioritization is crucial in projects that do not follow a strict policy of resolving only high and critical-severity findings. 

Furthermore, the \textit{Security Flama} has shown to be less transparent to its users than desirable. Especially during the first interviews, our subjects disagreed with the statement that the methodology was transparent in how data is processed. As shown in Figure~\ref{fig:trans}, this improved over time so that the average opinion of subjects was a slight agreement with the statement during the last interview. This is, however, not acceptable as an unclear functionality could lead to acceptance issues in the team. A subject in the final interview further requested a processing view of the data, depicting the impact of each processing stage. \\

Even though our evaluation identified different limitations of the \textit{Security Flama}, also several benefits were collected. 
The key benefit identified by our subjects was the evidence-based visibility over the current level of known security findings in the software product. The perceived benefits of this continuous security overview covered several perspectives of the data (e.g., new findings, most critical findings), reinforcing the importance of knowledge base queries, as defined in Subsection~\ref{subsec:automate}. Furthermore, they described this view as particularly useful for discussing security findings in team meetings. Hence, we believe that the enablement of team collaboration on security findings represents the ultimate benefit of our \textit{Security Flama}. 

Another benefit we derive from the evaluation is that it supported all process steps of the security findings management in the projects. Even though all subjects agreed with the statement that the \textit{Security Flama} covers all aspects of the security findings management in general, we believe that this claim cannot be generalized. In each project, different aspects of the security findings management are relevant. This creates a gap between claiming it was covering all aspects of the affected projects and covering all aspects in general. Therefore, we reduce the claim by concluding it was at least supporting all aspects of the selected projects. 

\begin{mdframed}[backgroundcolor=lightgray]
Benefits: Evidence-Based Overview, Collaboration Enablement, Support for all Management Activities\\
Limitations: Communication Strategy, Transparency of Methodology, Effort for Userinput
\end{mdframed}

\subsection{Threats to Validity}
\label{subsec:threats}
As with any evaluation of research results, also the validity of our conclusions is affected by several threats. 

First, the evaluation was conducted with the \textit{Security Flama}, an instance of the methodology. This presents a threat to the internal validity of the research, as problems in the implementation could be attributed to limitations of the methodology. Since automation, however, is the central part of the methodology, it is impossible to verify the impact without instantiating it. 

Another threat to the internal validity is the experience of our subjects with the management of security findings. During the evaluation, it was clear that some subjects were inexperienced in dealing with security findings and consequently associated advantages of the source data (suggested solution approaches) as benefits of the \textit{Security Flama}. Conversely, data of low semantic value was attributed as a limitation of the \textit{Security Flama}. Moreover, this also affects the construct validity of the interview data, as subjects with varying security expertise could interpret the questions differently. However, this represents practice realistically, as not every team member has expertise and experience in the security domain. 

The environment of our evaluation represents the threat to the external validity of our results. Since both projects are conducted in the same company, multiple factors affect whether the results can be transferred to other companies and projects. Especially the size of the project teams and the investigation of just two projects impact the validity of the results. However, we accept this threat since it is common with evaluations in industrial practice and consider a large scale evaluation as part of our future work. Moreover, we encourage researchers and industry practitioners to replicate our work with potentially different perspectives. For that purpose, we provide the introductory material for the evaluation, the interview guide, and both questionnaires in the supplementary material \cite{sup_material}. 
\subsection{Conclusion and Future Work}
\label{sec:conclude}
Managing security findings with the same efficiency as other DevOps practices challenges practitioners throughout the entire software engineering life cycle.
In this paper, we proposed a methodology for the management of security findings aligned with DevOps principles. To evaluate the impact on industrial practice, we created the \textit{Security Flama}, implementing our methodology as a semantic knowledge base for the management of security findings. The \textit{Security Flama} was integrated into two ongoing software development projects at a multinational industrial partner and its impact was evaluated with quantitative and qualitative methods. We conclude that both the methodology and its instance, the \textit{Security Flama}, establish the DevOps principle of continuous feedback \cite{Kimphoenix_2018} for security findings in software products while reinforcing other DevOps practices like cross-functional collaboration. 

Our research yielded three key results. 
First, the evaluation within the context of two industry projects shows that the usage of an automated methodology for the management of security findings is crucial for industrial DevOps projects. 
Second, the proposed methodology for security findings management is beneficial for the investigated projects. This was evidenced by interviews with the team members as well as represented by the impact on the project performance indicators. However, it was also noticed that we initially considered some aspects of the methodology to be more beneficial as it has proven to be (prioritization).
Third, the interview results reinforced the importance of a high-quality communication strategy and web interface. Most interview comments addressed limitations and improvements to the communication strategy, indicating its importance to our subjects. 

As part of the last result, the evaluation subjects provided multiple suggestions on how to improve the methodology. A redesign of the communication strategy therefore sets our future work on the methodology, including e.g. a reduction of effort to add user input and a deep integration of project backlogs. Since both projects wished to continue using the \textit{Security Flama}, the so-acquired long-term data will further complement the results of this paper in the future.

\begin{acks}
All authors would like to thank the members of both projects for the active collaboration on the topic and for the valuable feedback.
Furthermore, we would like to thank Berk Sudan and Sandip Sah for supporting the implementation of the Security Flama. 
\end{acks}

\bibliographystyle{ACM-Reference-Format}
\bibliography{literature}

\end{document}